# papaya2: 2D Irreducible Minkowski Tensor computation

**Fabian M. Schaller**[1, 2], **Jenny Wagner**[3], **and Sebastian C. Kapfer**[1]

**1** Theoretische Physik 1, FAU Erlangen-Nürnberg, Germany **2** Institut für Stochastik, Karlsruhe Institute for Technology, Germany **3** Zentrum für Astronomie, Universität Heidelberg, Germany

## Summary

A common challenge in scientific and technical domains is the quantitative description of geometries and shapes, e.g. in the analysis of microscope imagery or astronomical observation data. Frequently, it is desirable to go beyond scalar shape metrics such as porosity and surface to volume ratios because the samples are anisotropic or because direction-dependent quantities such as conductances or elasticity are of interest. Popular analysis software such as ImageJ and SExtractor provide only limited tooling for higher-order anisotropy characterization; usually only the tensor of inertia (rank 2) is available.

Minkowski Tensors are a systematic family of versatile and robust higher-order shape descriptors, originating in integral geometry, see Schröder-Turk et al. (2011) for an introduction and detailed references. They allow for shape characterization of arbitrary order and promise a path to systematic structure-function relationships for direction-dependent properties. Minkowski Tensors have previously been applied to data as diverse as ice grain microstructure (G. E. Schröder-Turk et al., 2010), granular packing geometries (Schaller et al., 2015; G. E. Schröder-Turk, Mickel, et al., 2010), astronomical data (Joby, Chingangbam, Ghosh, Ganesan, & Ravikumar, 2019; Kerscher et al., 2001; Klatt & Mecke, 2020), neuronal data (Beisbart, Barbosa, Wagner, & Costa, 2006), foams (Evans, Schröder-Turk, & Kraynik, 2017; Saadatfar et al., 2012) and random sets, tessellations and point patterns (Kapfer et al., 2010; Klatt et al., 2017). An accessible introduction to Minkowski Tensors can be found on www.morphometry.org.

Here, we present papaya2, a C++ library which facilitates computation of Irreducible Minkowski Tensors for two-dimensional geometries and shapes, including planar objects bounded by polygonal contours, collections of points (point patterns) and greyscale pixel data. This library is accompanied by example programs and bindings for Python, Matlab, and the JavaScript language.

Papaya2 is a rewrite of papaya with a library interface, support for Irreducible Minkowski Tensors and interpolated marching squares, and extensions to Matlab, JavaScript and Python provided. While the tensor of inertia is computed by many tools, we are not aware of other open-source software which provides higher-rank shape characterization in 2D.

For the analysis of the examples in this paper, we employ our interactive online resource Morphometer which uses papaya2 for its computations.

## C++ library papaya2

The C++ 11 library papaya2 contains the core algorithms to compute Irreducible Minkowski Tensors of two-dimensional geometries. It processes both polygonal and 2D image input data.

papaya2 is a header-only template library designed to operate on user data structures. We bundle several example programs which can be adapted to user requirements, or employed directly for simple analyses (see section *Demos*).

The main components of the library are defined in the header file `<papaya2.hpp>`. Analysis results are returned in a `MinkowskiAccumulator` object, which offers accessors to retrieve common morphometric data, including the following:

- `area()` The 2D volume (area) enclosed by the geometry
- `perimeter()` The perimeter (boundary length) of the geometry
- `msm(s)` The $s$-th Minkowski structure metric $q_s$, see [Morphometry page](#) and Mickel, Kapfer, Schröder-Turk, & Mecke ([2013](#)) for details
- `imt(s)` The $s$-th Irreducible Minkowski Tensor $\Psi_s$, see previous item for details

The library provides convenient wrapper functions which encapsulate common analysis tasks. In general, these functions are C++ function templates which operate on user data structures. User-supplied data structures need to include some required methods and operators as documented in the headers. The most important entrypoints are

- `papaya2::imt_polygon`: compute the Irreducible Minkowski Tensors of closed simple polygons, specified as a sequence of vertices in counterclockwise order.
- `papaya2::imt_interpolated_marching_squares`: computes the Irreducible Minkowski Tensors of an excursion set of a single channel of a raster graphics image (bitmap). An extended version of the Marching Squares algorithm is used which computes interpolated contours from 2x2 neighborhoods, see Mantz, Jacobs, & Mecke ([2008](#)) for details. The input data is passed to papaya2 by reference via a suitable adapter class to avoid copies. There are several examples of adapter classes provided, as well as a copying container (`BasicPhoto`).
- `papaya2::minkowski_map_interpolated_marching_squares`: implements the Minkowski map algorithm (Schröder-Turk et al., [2010](#)) for a space-resolved anisotropy analysis.

The supplementary header `<papaya2/voronoi.hpp>` implements the Minkowski Tensor analysis of point patterns via the Voronoi tessellation approach (Kapfer et al., [2010](#)). The demo `ppanalysis` exemplifies how to use this header file.

## Application Examples

Here we show some examples analyzed in the [Morphometer web application](#), which uses `papaya2.js`, the JavaScript version of the `papaya2`. Morphometer provides rapid analysis of small amounts of data (up to 1000 points, or 500x500 pixels). For routine analysis we recommend using the `ppanalysis` and `imganalysis` demos or Python/Matlab bindings.

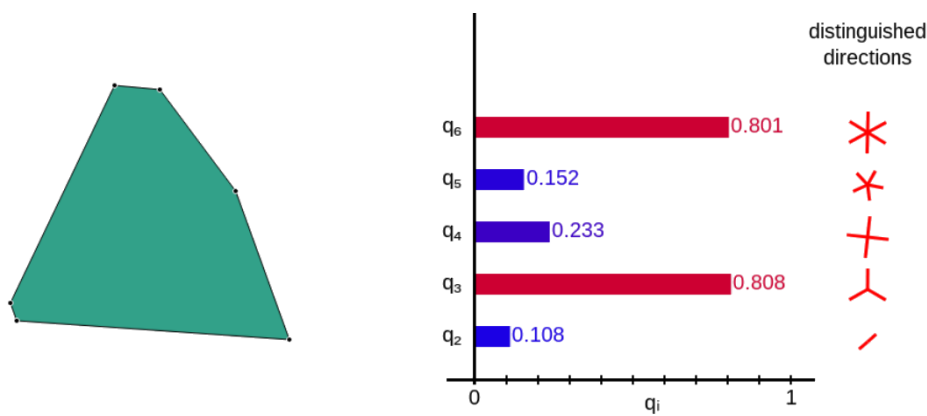

**Figure 1:** Minkowski Tensor analysis of a polygon in Morphometer.

Minkowski Tensors can be applied to different types of data:

- Single polygons: $s$-fold symmetric polygons are characterized by high values of $q_s$. Figure 1 shows a polygon with approximates an equilateral triangle. Therefore, we find high values of $q_3$, $q_6$, $q_9$, etc. The distinguished directions of each $\Psi_s$ are depicted on the right of the $q_s$ bar diagram.

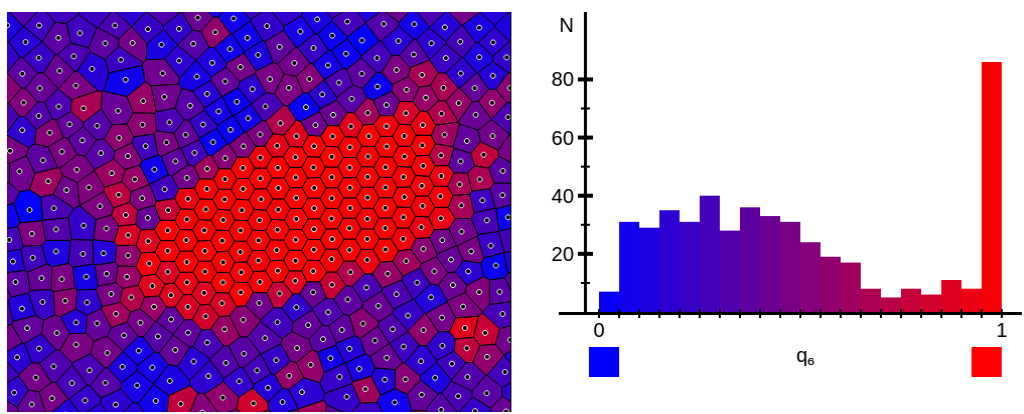

**Figure 2:** Minkowski Tensor analysis of a point pattern induced by a granular crystal cluster.

- Point patterns can be, for instance, realizations of abstract point processes or data of physical particle systems. For the Minkowski Tensor analysis, a Voronoi tessellation of the points is constructed and Minkowski Tensors of the individual Voronoi cells are computed. Figure 2 (left) shows a hexagonal crystal cluster surrounded by an amorphous background. The Minkowski structure metric $q_6$ (indicated by the color of the Voronoi cells) is very well suited to detect hexagonal crystalline structures. The presence of ideal hexagonal cells is demonstrated by the peak at $q_6 = 1$ in the histogram on the right-hand side.

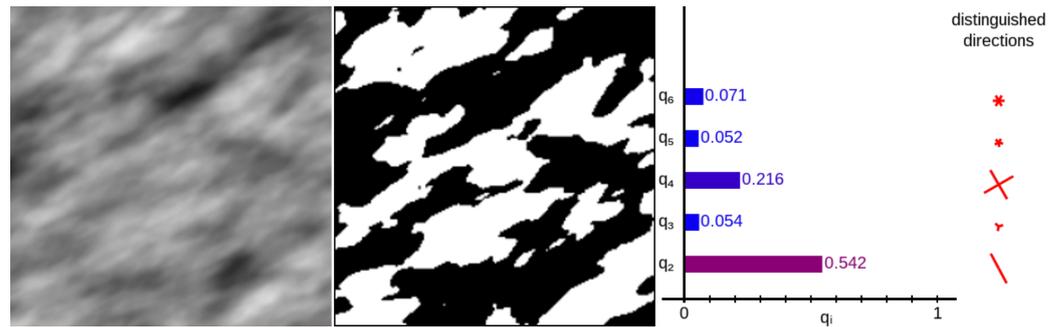

**Figure 3:** Minkowski Tensor analysis of a greyscale image: a Gaussian random field.

- Greyscale images can also be analyzed in terms of Minkowski Tensors. Figure 3 (left) shows a detail of an anisotropic Gaussian random field, which is converted into a binary image by thresholding (center) and analyzed using Minkowski Tensors (right). The significant $q_2$ value in the Minkowski analysis (right) shows that the random field has a preferred direction, which is also reflected by the distinguished direction marker (red color).

## Demos and language bindings

In the directory demos, we provide a number of example programs which use the library for data analysis. These are meant to be modified and adapted to user needs as required. For simple analyses, they can be used directly, see the README file in the demos folder.

We also provide bindings of the library for Python, Matlab and JavaScript.

## Acknowledgements

We acknowledge funding by Deutsche Forschungsgemeinschaft as part of the Forschergruppe GPSRS.

We are grateful to Daniel Hug, Günther Last, Klaus Mecke and Gerd Schröder-Turk for guidance and support, to Michael Klatt for many discussions and for his contributions to the scientific concept and content of the Morphometer, to Dennis Müller and Thomas Schindler for example data, and to Simon Weis for technical advice.

We use picopng for loading PNG images, emscripten for compiling to JavaScript, CGAL for Voronoi diagrams, and Catch2 for unit tests.